\documentclass[12pt,openany]{cmuthesis}

\usepackage{times}
\usepackage{fullpage}
\usepackage{graphicx}
\usepackage{amsmath}
\usepackage[numbers,sort]{natbib}
\usepackage[backref,pageanchor=true,plainpages=false, pdfpagelabels, bookmarks,bookmarksnumbered,
]{hyperref}
\usepackage{subcaption}
\usepackage{url}
\usepackage{booktabs}
\usepackage{tabularx}
\usepackage{bbm}
\usepackage{xcolor}
\usepackage{amsmath}
\usepackage{resizegather}
\usepackage{enumitem}
\usepackage{array,multirow}
\usepackage{soul}

\usepackage{soul,xcolor}

\usepackage[letterpaper,twoside,vscale=.8,hscale=.75,nomarginpar,hmarginratio=1:1]{geometry}

\begin {document} 
\frontmatter

%initialize page style, so contents come out right (see bot) -mjz
\pagestyle{empty}

\title{ %% {\it \huge Thesis Proposal}\\
{\bf Expediting Support for Social Learning with Behavior Modeling}}
\author{Yohan Jo, Gaurav Tomar, Oliver Ferschke, Carolyn P. Ros\'{e}, Dragan Ga\v{s}evi\'{c}}
\date{}
\Year{2016}
\trnumber{CMU-LTI-16-011}

% \committee{
% }

% \support{}
% \disclaimer{}

% copyright notice generated automatically from Year and author.
% permission added if \permission{} given.

%\keywords{Stuff, More Stuff}

\maketitle

% \begin{dedication}
% For my dog
% \end{dedication}

%\pagestyle{plain} % for toc, was empty

%% Obviously, it's probably a good idea to break the various sections of your thesis
%% into different files and input them into this file...

\begin{center}
%\begin{minipage}[t]{4.875in}   % must be a minipage due to \thanks
\begin{minipage}[t]{4.8in}      % hackery
\vbox to 2in{
\vfill
\begin{center}
\vskip 4em
{\Large \bf Expediting Support for Social Learning with Behavior Modeling}
\vskip 3em
{\large Yohan Jo, Gaurav Tomar, Oliver Ferschke, Carolyn P. Ros\'{e}}
\vskip .5em
{\sc
 School of Computer Science\\Carnegie Mellon University\\ Pittsburgh, PA, USA\\
 \{yohanj, gtomar, ferschke, cprose\}@cs.cmu.edu
}
\vskip 1em
{\large Dragan Ga\v{s}evi\'{c}}
\vskip .5em
{\sc
 Schools of Education and Informatics\\The University of Edinburgh\\Edinburgh, UK\\
 dgasevic@acm.org
}
\end{center}
\vfill}
\end{minipage}
\end{center}

\begin{abstract}
An important research problem for Educational Data Mining is to expedite the cycle of data leading to the analysis of student learning processes and the improvement of support for those processes. For this goal in the context of social interaction in learning, we propose a three-part pipeline that includes data infrastructure, learning process analysis with behavior modeling, and intervention for support. We also describe an application of the pipeline to data from a social learning platform to investigate appropriate goal-setting behavior as a qualification of role models. Students following appropriate goal setters persisted longer in the course, showed increased engagement in hands-on course activities, and were more likely to review previously covered materials as they continued through the course. To foster this beneficial social interaction among students, we propose a social recommender system and show potential for assisting students in interacting with qualified goal setters as role models. We discuss how this generalizable pipeline can be adapted for other support needs in online learning settings.
\end{abstract}

% \begin{acknowledgments}
% My advisor is cool.
% \end{acknowledgments}

% \tableofcontents
% \listoffigures
% \listoftables

\mainmatter

%% Double space document for easy review:
%\renewcommand{\baselinestretch}{1.66}\normalsize

% The other requirements Catherine has:
%
%  - avoid large margins.  She wants the thesis to use fewer pages, 
%    especially if it requires colour printing.
%
%  - The thesis should be formatted for double-sided printing.  This
%    means that all chapters, acknowledgements, table of contents, etc.
%    should start on odd numbered (right facing) pages.
%
%  - You need to use the department standard tech report title page.  I
%    have tried to ensure that the title page here conforms to this
%    standard.
%
%  - Use a nice serif font, such as Times Roman.  Sans serif looks bad.
%
% Other than that, just make it look good...

\chapter{Introduction}
\label{sec:introduction}

More and more recent work in educational data mining and learning analytics refers to a ``virtuous cycle'' of data leading to insight on what students need and then improvements in support for learning~\cite{Thille2010}. An important goal is tightening this cycle to improve learning experience. We are interested especially in social learning, drawing from a Vygotskian theoretical frame where learning practices begin within a social space and become internalized through social interaction. This may involve limited interaction, such as observation, or more intensive interaction through feedback, help exchange, sharing of resources, and discussion.

There are two main contributions of this paper. The first is to propose a pipeline and its component models that can expedite the cycle of data mining technology used to make sense of pathways of learner behaviors. The second is to present findings from an application of the proposed pipeline for the purpose of addressing a specific problem in goal-setting in a social learning platform.

\begin{figure}
    \centering
    \includegraphics[width=\linewidth]{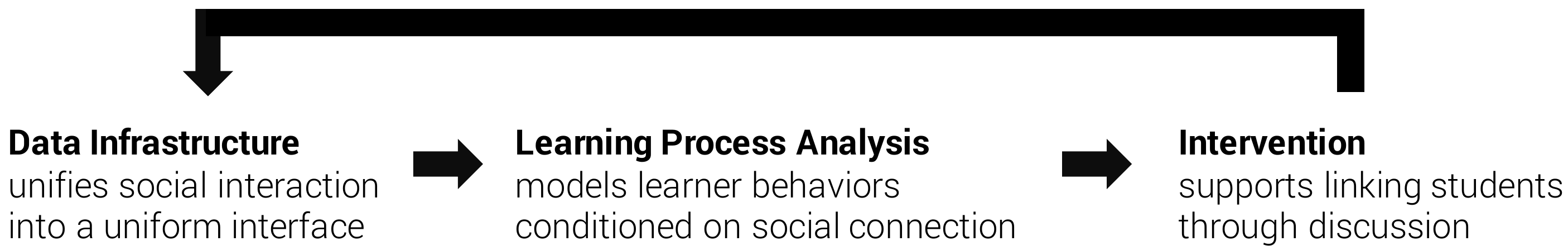}
    \caption{Pipeline for educational data mining in social learning.}
    \label{fig:pipeline}
\end{figure}

Specifically, our first contribution is to propose a pipeline that can expedite the cycle of data infrastructure, learning process analysis, and intervention (Figure~\ref{fig:pipeline}). Data infrastructure provides a uniform interface for heterogeneous data from social interaction in various platforms, such as connectivist Massive Open Online Courses (cMOOCs)~\cite{siemens2014connectivism}, hobby communities, and Reddit communities, where people engage in follower-followee relations, post updates to their account, engage in threaded discussions, and also optionally link in blogs, YouTube videos, and other websites. Learning process analysis aims to analyze students' processes depending on their social network configurations and to identify beneficial kinds of social connections. We developed a probabilistic graphical model that analyzes sequences of behaviors in terms of topics expressed and social media types that students actively engage in over time. Finally, intervention is introduced to foster beneficial social connections among students. We developed a recommender system that matches qualified students to discussions to increase opportunities for them to interact with other peers. The pipeline is iterative such that data from participation is used to create models that trigger interventions in subsequent runs of the course. Data from those later runs can be used to train new and better models in order to improve the interventions, and so on.

Our second contribution is to present findings from an application of the proposed pipeline to data from a social learning environment called ProSolo~\cite{Rose2015}, in order to investigate the positive influence of observing goal-setting behavior. While goal-setting has been intensively researched and proven to be an important self-regulated learning (SRL) practice that often leads to success in learning, the influence of a student's goal-setting behavior on observers has little been investigated empirically. If goal-setting students turn out to be good role models, that is, beneficial to their social peers, we can encourage and help students to make such social connections with goal setters to enhance their learning experience. The usefulness of this effect may be especially desirable in online courses where the number of instructors is limited, or online communities that are not structured like courses, where students are required to take more agency in forging a learning path for themselves within an ecology of resources.

In the remainder of this paper, we begin by motivating the specifics of our pipeline as situated within the literature. Next, we present our pipeline and its application, along with concrete computational models and findings.

\chapter{Related Work}
\label{sec:related_work}

We first explore the literature on social learning and peer effects, which are the main context and motivation of our work. Next, we relate the components of our pipeline to related prior work.

\section{Social Effects On Learning}
Vygotsky's view of social interaction as a key to learning and Bandura's social learning theory~\cite{Bandura1971} emphasize the importance of interaction to learning. In social contexts, by vicarious learning, students observe external models and learn from those observations even when not actively engaged in interaction~\cite{Winne2010}. Observation of role models facilitates motivation and self-efficacy for a task~\cite{Schunk1985} and may be associated with positive changes in the observer's behavior~\cite{Paluck:2016fc}. Drawing on this theoretical foundation, the positive impact of social interaction has been investigated in collaborative work~\cite{Molenaar2015} and in online courses~\cite{Rose2014}. Yet, to our knowledge, our work is the first to investigate goal-setting behavior specifically as a qualification of a role model in online learning.

\section{Data Infrastructures}
Several data infrastructures have been introduced to aid educational data mining for Massive Open Online Courses (MOOCs). For instance, MOOCdb \cite{Veeramachianeni2014} and DataStage\footnote{\url{http://datastage.stanford.edu/}}, designed to store raw data from MOOCs, consolidate clickstream data from different MOOC platforms in a single, standardized database schema. This allows for developing platform-independent analysis tools, thus enabling analyses that span multiple courses hosted by different MOOC providers with reduced development effort. While these infrastructures focus on behavior data represented by clickstream logs, our proposed infrastructure deeply represents other aspects of student interactions, such as discussion behavior and social relationships, which require the natural language exchange between students.

DataShop~\cite{Koedinger2013a} is another repository of learning data that focuses on the interaction between students and educational software. DataShop offers a set of tools for analyzing these datasets and building cognitive models that allow researchers to explore the relationships between students' skills, concepts or misconceptions, and their trajectory in learning environments. Compared to human-computer interaction, however, analysis of human-human interaction is more complex, where interaction states are implicit, continuous, and controlled by a decentralized structure. Furthermore, in social learning, it is important to consider  social relationships between humans and how those relationships moderate the effect of any computer agent or learning platform involved. 

%Other large scale repositories include Databrary~\cite{Databrary2015} for video data in the developmental sciences and TalkBank~\cite{TalkBank2004}, which is a collection of databases for the research of human and animal communication.

\section{Learning Process Analysis}

Analysis of students' learning processes has been a critical topic in education. Our method contributes to the literature on process mining through behavior modeling. Approaches to learning process analysis differ in the definition of the basic building block, often conceived of as states within a graph. Common building blocks for tutoring systems and educational games include knowledge components~\cite{Zhao2006} and actions~\cite{Rowe2015}. In dialogue settings, it is common to code each utterance according to a coding scheme and analyze the sequence of codes~\cite{Ezen-Can2015,Molenaar2015}. In a MOOC context, states are often defined as course units~\cite{Coffrin2014,Kizilcec2013}, course materials~\cite{Coffrin2014}, and discussion threads~\cite{Bogarin2015}. 

Such predefined states, however, may not be the ideal units of states, especially in online courses where students can selectively engage in learning resources. Therefore, unsupervised modeling approaches are appealing for the purpose of identifying states that are meaningful indications of student interests obtained in a data-driven way. Markov models have been proposed to learn latent states and state transitions~\cite{Yang2014,Tang2012}. However, their representation of a state is often too simple, e.g., a single multinomial distribution over observations. To improve the simple representation, the state transition topic model has been proposed~\cite{Jo2015}, in which a state is represented as a mixture of topics, from which documents are generated via Latent Dirichlet Allocation (LDA). Yet, this type of model does not consider conditional state transitions, thereby imposing limitations in modeling the complex dynamics of social learning environments. Our model extends this basic approach so as to incorporate more information related to our interest, by distinguishing between multiple document types and conditioning state transitions on different types of social connection.

While we learn states based on the topics of student discussions in a course, there are also other views on the use of discussions for defining learning process. Milligan~\cite{Milligan2015} argues that participation in discussion does not necessarily mean genuine learning of the content, so we should consider more latent and complex learning skills reflected in the content. Ezen-Can et al.~\cite{Ezen-Can2015} go deeper into discussion text to analyze the cognitive process evidenced in the discussion rather than the mere topics of the content.

\section{Social Recommendation}

In MOOCs, a student's learning process is affected by other peers especially through interaction in forums, which offer opportunities to develop communication and community. %However, in large-scale courses, forums commonly lose participation due to poor thread management and an overwhelming number of discussion threads \cite{mackness2010ideals}. When the forums fail to properly sustain a sense of community, high rates of student dropout often follow. 
Hence, social recommendation algorithms can introduce appropriate students to certain discussions for productive interaction. Suggested matches should be appropriate when viewed from both discussion and student sides \cite{terveen2005social}, for example by suggesting a student to participate in discussions based on both the potential benefit of the student's expertise as an asset to the discussions while respecting the limitations of a student's resources for participation in more than a limited number of discussions~\cite{yang2014question}. Our model can recommend discussions to a student by balancing the benefit of the student's qualification to discussions, her relevance to discussions, and required effort.

\chapter{Three-Part Analytics Pipeline}
\label{sec:pipeline}
Our pipeline is designed to expedite the process of exploiting student data leading to data-driven decision-making for enhancing student learning (Figure~\ref{fig:pipeline}).

In this pipeline for social learning, the first component is a data infrastructure that maps diverse forms of social interaction into a common structure. This uniform interface allows the subsequent components---learning process analysis and intervention---to apply the same tools to different data, even from distinctly different discourse types, with little modification. Our development of this infrastructure, DiscourseDB\footnote{\url{http://discoursedb.github.io}}, represents discourse-centered social interaction as an entity-relation model. Discourses (e.g., forums or social media) and individual contributions in a discourse (e.g., posts, comments, and utterances) are represented as generic containers generalizable to diverse social platforms. DiscourseDB also allows for defining arbitrary relations between contributions, e.g., a ``reply-to'' relation derived from the explicit reply structure of the platform versus one inferred through some automated analysis process. This flexibility helps the subsequent components of the pipeline avoid data-specific processing. DiscourseDB can store both active and passive activities of individuals, such as creating, revising, accessing, and following contributions, as well as forming social connections with other individuals. DiscourseDB is the key component of our pipeline, based on which the next components perform integrated analyses of discourses and social networking on multiple platforms with reusability.

The second component of our pipeline is analysis of students' learning processes depending on their social connections. The goal is to assess students' needs of support by understanding how learning processes are affected by social interaction and what types of social interaction are helpful to students. Just as Bayesian knowledge tracing enables modeling the learning process from a cognitive perspective and then supporting a student's progress through a curriculum, Bayesian approaches can model learning processes at other levels, including supportive social processes. And similarly, these models can then be used to trigger support for the learning processes in productive ways. Hence, the third component of our pipeline draws upon insights obtained from the analysis to introduce interventions that can help students make beneficial social connections with other peers. We will propose two concrete examples of machine learning techniques for these two components in Section \ref{sec:assessment} and Section \ref{sec:support} respectively.

\chapter{Research Context}\label{sec:research_context}

The remainder of the paper presents an example application of our general pipeline to a specific problem. This section describes the problem of our interest, data set, and terminology that will be used throughout the application.

\section{Problem and Data} 
We examine goal-setting behavior as a potential qualification of good role models via learning process analysis and foster social connections with goal setters via recommendation support. Since most MOOCs and informal learning communities lack a measure to identify potentially good role models (e.g., a pretest), increased frequency of effective goal-setting behaviors may serve as an indirect indicator of success, as previous studies showed positive relationships between goal-setting behavior and learning outcomes \cite{Battle2003,Husman2008,Zimmerman2008}.

The data was collected from an edX MOOC entitled \emph{Data, Analytics, and Learning} (DALMOOC)~\cite{Rose2015}, which ran from October to December 2014. This course covered theoretical principles about learning analytics as well as tutorials on social network analysis, text mining, and data visualization. This MOOC was termed a \emph{dual layer} MOOC because students had the option of choosing a more standard path through the course within the edX platform or to follow a more self-regulated and social path in an external environment called ProSolo. The ProSolo layer allowed students to set their own learning goals and follow other students so that they could view activities and documents that offered clues about how to approach the course productively. While a huge literature on analysis of MOOC data focuses on Coursera, edX, and Udacity MOOCs, other platforms with more social affordances are growing in popularity. In order to serve the goal of identifying support needs and automating support that may be triggered in a social context, it is advantageous to work with data from socially-oriented platforms. We used the log data from ProSolo as our object of analysis, which include students' discussions on ProSolo and their own blogs and Twitter that they identified on their ProSolo profile pages, evidence of students' social connection with each other, and ``goal notes,'' which students can use to set their learning goals in their own words.

We preprocessed discussion data before running our model. First, we filtered course-relevant tweets using the hashtags \#prosolo, \#dalmooc, and \#learninganalytics. We confirmed that the tweets identified as irrelevant by this process have little to do with course activity. Because we are not interested in irrelevant content, we replaced such content with a tag to indicate irrelevant content. In order to prevent topics from being defined in terms of document types, we removed Twitter mentions and ``RT'' from tweets as well as other function words including URLs from all documents. Descriptive statistics for the data set are listed in Table~\ref{tab:prosolo_stats}.

\section{Goal Quality and Social Connection}
To categorize the quality of goal-setting behavior of each student, we first annotated each goal note written by students indicating whether it indeed contains a goal or not. 58\% of goal notes contained goals. An example goal note is as follows: \emph{``to understand learning analytics and see how these may be useful for my teaching and in particular, my learning resource design/development.''} On the basis of this annotation, we categorized students into three classes: (1) goal setters, (2) goal participants, and (3) goal bystanders. Goal setters have goal notes that mention their distal or/and proximal goals. Goal participants have goal notes, all of which are about something other than goals, e.g., experiences or questions. Goal bystanders have no goal notes. Note that the category of a student can change over time. All students start as goal bystanders and may become a goal participant or a goal setter as time passes. A student's \emph{social connection} is then categorized into seven classes: (S1) has already been following a goal setter, (S2) started to follow a goal setter at the current time point (S3) has been following a goal participant (but no goal setter), (S4) started to follow a goal participant at the current time point, (S5) has been following a goal bystander (at best), (S6) started to follow a goal bystander at the current time point, and (S7) follows no one. S2, S4, and S6 mean that a student's social connection improved at the current time point, whereas S1, S3, and S5 indicate that a student remained in the same social connection category as in the previous time point.

\begin{table}[tbp]
\small
\centering
\begin{tabularx}{\columnwidth}{ p{40mm}X p{40mm} X }
\toprule
Goal notes& 62 & Tweets (relevant)& 715\\
ProSolo posts& 318 & Tweets (irrelevant)& 25,461\\
Blog posts& 359 & &\\ \midrule
Users & 1,729 & Social connections & 814 \\
\bottomrule
\end{tabularx}
\caption{Descriptive statistics for ProSolo data.}
\label{tab:prosolo_stats}
\end{table}

\chapter{Learning Process Analysis}\label{sec:assessment}

The second component of the pipeline aims to assess students' needs of support. Hence, we model students' behavior and analyze their learning processes especially as they experience changes in their social connections over time in the course. What models are best to use depends on the specific analysis of interest. In this section, we propose a Bayesian model designed for our problem and present our findings.  In particular, we found out that a student's learning process is positively related to her social connections with goal setters.  This pattern suggests the potential positive impact of supporting interaction with goal setters.

\section{Model}

Our model automatically extracts a representation of students' learning processes based on their discussions in a course and their social connections, which may reveal the influence of different configurations within the social space. We define the building blocks of learning processes, which we call states, in terms of discussed topics and the document types used for discussions (e.g. Twitter, blog). Given students' sequences of documents and social connection types over time, the model infers a set of meaningful states, along with the topics and document types for each state. The learned topics provide the information about students' interests, and the document types give an insight into how students use different media for different interests. The model also learns transition probabilities between states, conditioned on the category of a student's social connection in the source state.

\begin{figure}[tp]
    \small
    \begin{enumerate}
    	\item For each topic $j=0, ..., Z-1$,
    	\begin{enumerate}
    		\item Draw a word distribution $\phi_j \sim \text{Dirichlet}(\beta)$
    	\end{enumerate}
    	\item For each state $c=0, ..., S-1$,
    	\begin{enumerate}
    		\item Draw a document type distribution $\psi_c \sim \text{Dirichlet}(\nu)$
    		\item Draw a topic distribution $\theta_c \sim \text{Dirichlet}(\alpha)$
    		\item For each category of social connection $b=0, ..., A$,
    		\begin{enumerate}
        		\item Draw a transition distribution $\pi_{cb} \sim \text{Dirichlet}(\gamma)$
        	\end{enumerate}
    	\end{enumerate}
    	\item For each time point $t=1, 2, ...$,
    	\begin{enumerate}
    	    \item Choose a state $s_t \sim \text{Categorical}(\pi_{s_{t-1}})$
        	\item For each document,
        	\begin{enumerate}
        		\item Choose a document type $d \sim \text{Categorial}(\psi_{s_t})$
        		\item For each word,
        		\begin{enumerate}
        			\item Choose a topic $z \sim \text{Categorical}(\theta_{s_t})$
        			\item Choose a word $w \sim \text{Categorical}(\phi_z)$
        		\end{enumerate}
        	\end{enumerate}
        \end{enumerate}
    \end{enumerate}
    \caption{Generative process of our model.}
    \label{fig:sttm+_process}
\end{figure}

\begin{figure}[tp]
    \centering
    \includegraphics[width=.7\linewidth]{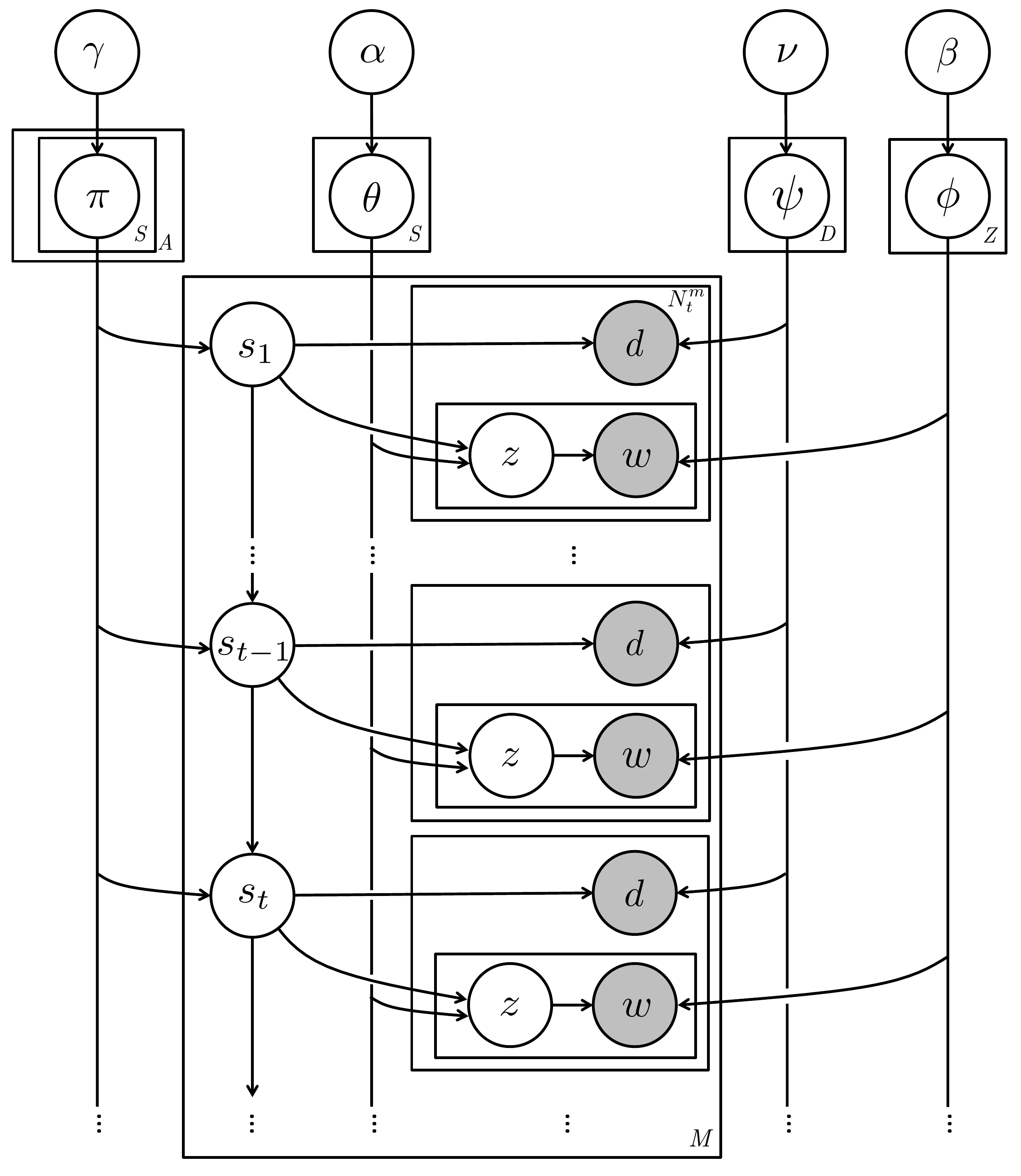}
    \caption{Graphical representation of our model.}
    \label{fig:sttm+}
\end{figure}

\begin{table}[tp]
    \centering
    \bgroup
    %  1 is the default, change whatever you need
    \small
    \begin{tabularx}{\columnwidth}{cX} 
    \toprule
    $S$ & number of states \\
    $A$ & number of social category types \\
    $Z$ & number of topics \\
    $D$ & number of document types \\
    $M$ & number of sequences \\
    ${\sf docs}^m_t$ & documents at $t$-th time point in sequence $m$  \\
    $w^m_{t,i}$ & $i$-th word in ${\sf docs}^m_t$  \\
    $z^m_{t,i}$ & topic assigned to $w^m_{t,i}$ \\
    $s^m_t$, $\bar{\textbf{s}}^m_t$ & state at $t$-th time point in sequence $m$  \\
    $a^m_t$ & social status at $t$-th time point in sequence $m$  \\
    $\textbf{w}^m_t$, $\bar{\textbf{w}}^m_t$ & all words in ${\sf docs}^m_t$, $\textbf{w} - \{\textbf{w}^m_t\}$  \\
    $\textbf{z}^m_t$, $\bar{\textbf{z}}^m_t$ & topics assigned to $\textbf{w}^m_t$, $\textbf{z} - \{\textbf{z}^m_t\}$  \\
    $\textbf{d}^m_t$ & document types of ${\sf docs}^m_t$  \\
    $N^{MTZ}_{m,t,j}$ & number of words assigned topic $j$ in ${\sf docs}^m_t$  \\
    $N^{ZW}_{j,w}$ & \# words $w$ assigned topic $j$  \\
    $N^{SD}_{c,k}$ & \# occurrences of document type $d$ in state $c$ \\
    $N^{SZ}_{c,j}$ & \# words assigned topic $j$ in state $c$ \\
    $N^{SAS}_{c,a,c'}$ & \# transitions from state $c$ to $c'$ given  status $a$  \\
    \bottomrule
    \end{tabularx}
    \egroup
    \caption{Description of notation.}
    \label{tab:sttm+_notations}
\end{table}

We extend the state transition topic model from prior work~\cite{Jo2015} by including components for document types and conditional state transitions. More formally, our generative model assumes that each state has probability distributions over document types and over topics. Each state also has a probability distribution over states for each type of social connection, representing transition probabilities to states (including itself). This model assumes a generative process of students visiting states and making documents as follows. Suppose that there is a set of states that constitute students' learning processes. At each time point, a student enters into a state and writes documents; he chooses a document type and then repeats selecting a topic and writing a word. At the next time point, the student enters into a state according to the state transition probabilities of the current state, which is conditioned on the student's current social connection. This generative process is described more formally in Figure~\ref{fig:sttm+_process}, and its graphical representation is shown in Figure~\ref{fig:sttm+}. Notations are explained in Table~\ref{tab:sttm+_notations}. 

Through an inference step, we can estimate the distributions over topics, document types, and transitions for each state ($\theta_c$, $\psi_c$, and $\pi_{cb}$, respectively). So also would the topics ($\phi$) and the state of each time point for students  ($s_t$) be estimated. We use Gibbs sampling for inference. Each iteration samples $z^m_{t,i}$ and $s^m_t$ according to the following probabilities.
%\resizebox{\columnwidth}{!}{
%   \begin{minipage}{\linewidth}
  \begin{align*}
    &p(z_{t,i}^m = j | \boldsymbol{\bar{z}}_{t,i}^m , \boldsymbol{{w}} , s_t^m) \propto  \left( N_{m,t,j}^{MTZ} + \alpha \right) \frac{N_{j,w_{t,i}^m}^{ZW} + \beta}{\sum_{w'} \left( N_{j,w'}^{ZW} + \beta \right)}, \\
    &p(s_t^m = c | \boldsymbol{\bar{s}}_t^m , \boldsymbol{z}, \boldsymbol{a}, \boldsymbol{d}) \propto  \left(\prod_{k=1}^{D} \frac{\Gamma\left(N^{SD}_{c,k}+\nu + N^{MTD}_{m,t,k}\right)}{\Gamma\left(N^{SD}_{c,k}+\nu\right)}  \frac{\Gamma\left(\sum_{k'}(N^{SD}_{c,k'}+\nu)\right)}{\Gamma\left(\sum_{k'}(N^{SD}_{c,k'}+\nu)+|\boldsymbol{d}^m_t | \right)}  \right) \\ 
    & ~~~~~~~~~~\times \left(\prod_{j=1}^{Z} \frac{\Gamma\left(N^{SZ}_{c,j}+\alpha + N^{MTZ}_{m,t,j}\right)}{\Gamma\left(N^{SZ}_{c,j}+\alpha\right)}  \frac{\Gamma\left(\sum_{j'}(N^{SZ}_{c,j'}+\alpha)\right)}{\Gamma\left(\sum_{j'}(N^{SZ}_{c,j'}+\alpha)+|\boldsymbol{z}^m_t | \right)}  \right) \\ 
    & ~~~~~~~~~~\times \left( \frac{N^{SAS}_{s^m_{t-1},a^m_{t-1},c} + \gamma}{\sum_{c'} \left( N^{SAS}_{s^m_{t-1},a^m_{t-1},c'} + \gamma \right)} \right) \\
    & ~~~~~~~~~~\times \left( \frac{N^{SAS}_{c,a^m_{t},s^m_{t+1}} + \boldsymbol{1}(s^m_{t-1}=c=s^m_{t+1})+\gamma}{\sum_{c'}\left(N^{SAS}_{c',a^m_{t},s^m_{t+1}} + \boldsymbol{1}(s^m_{t-1}=c'=s^m_{t+1})+\gamma\right) } \right).
\end{align*}
 % \end{minipage}
%}

From the sampling results, we can estimate 

\begin{align*}
	&\phi_{j,w} = \frac{N^{ZW}_{j,w} + \beta}{\sum_{w'}\left(N^{ZW}_{j,w'} + \beta \right)},
	\theta_{c,j} = \frac{N^{SZ}_{c,j} + \alpha}{\sum_{j'} \left(N^{SZ}_{c,j'} + \alpha \right)},\\
	& \psi_{c,k} = \frac{N^{SD}_{c,k} + \nu}{\sum{k'} \left( N^{SD}_{c,k'} + \nu \right)},
	\theta^m_{t,j} = \frac{N^{MTZ}_{m,t,j} + \alpha}{\sum_{j'} \left(N^{MTZ}_{m,t,j'} + \alpha \right)},\\
	& \pi_{c,b,c'} = \frac{N^{SAS}_{c,b,c'} + \gamma}{\sum_{c'} \left(N^{SAS}_{c,a,c'} + \gamma \right)},
\end{align*}
where $\theta^m_{t}$ is the topic distribution of ${\sf docs}^m_t$. A detailed derivation process and the source code are available on our website\footnote{\url{http://cs.cmu.edu/~yohanj/research}}. We can also infer the state of each time point of a student, e.g., by the state assigned to each time point during the sampling process. For an unseen sequence of documents, we may infer the state of each time point using a Viterbi algorithm based on the document type distribution, topic distribution, and state transition distribution of each state. Once states are finalized, maximum likelihood estimation can be applied to estimate the topic distribution of the documents in each time point.

%\footnote{\scriptsize\url{http://cs.cmu.edu/~yohanj/research/EDM16}}.

\section{Findings}

We applied the model to the ProSolo data and examined the correlation between the categories of social connection and learning behaviors. We ran our model with the number of states set to 10 and the number of topics set to 20. We defined the unit of a time point as one week, and if a student had no activity in a certain week, that week was omitted from her sequence.

\begin{table*}[t]
    \centering
    \small
    \begin{tabularx}{\textwidth}{ lX cccccc }
        \toprule
%        & & \multicolumn{6}{ c }{Document Types} \\ \cmidrule{3-8}
        State & Topics & RelGoalNote & IrGoalNote  & Post & Blog &     Rel\-Tweet & Ir\-Tweet \\ \midrule
        0 & Course-irrelevant tweets & 0.00 & 0.00 & 0.00 & 0.00 & 0.00 & 1.00 \\ 
        1 & Concept map, network analysis (Week 9) & 0.00 & 0.00 & 0.02 & 0.01 & 0.18 & 0.78 \\ 
        2 & Social capital (Week 3) & 0.04 & 0.01 & 0.19 & 0.30 & 0.18 & 0.27 \\ 
        3 & Tableau (Week 2), Gephi (Week 3), Lightside (Week 7) & 0.01 & 0.03 & 0.10 & 0.28 & 0.24 & 0.34 \\
        4 & Prediction models (Week 5) & 0.01 & 0.02 & 0.29 & 0.22 & 0.10 & 0.36 \\ 
        5 & Data wrangling (Week 2) & 0.01 & 0.01 & 0.12 & 0.08 & 0.26 & 0.52 \\
        6 & Visualization (Week 3) & 0.05 & 0.02 & 0.24 & 0.47 & 0.08 & 0.15 \\ 
        7 & Epistemology, assessment, pedagogy (Week 4) & 0.05 & 0.00 & 0.18 & 0.22 & 0.30 & 0.25 \\
        8 & Prediction, decision trees (Week 5) & 0.02 & 0.02 & 0.19 & 0.40 & 0.09 & 0.28 \\ 
        9 & Share, creativity (mixed topics) & 0.00 & 0.02 & 0.12 & 0.13 & 0.21 & 0.52 \\
        \bottomrule
    \end{tabularx}
    \caption{Learned states with their topics and document type distribution (each row sums to 1). ({\sf RelGoalNote}: goal notes containing a goal, {\sf IrGoalNote}: goal notes without a goal, {\sf Post}: posts on ProSolo, {\sf Blog}: personal blog posts, {\sf Rel\-Tweet}: course-relevant tweets, {\sf Ir\-Tweet}: course-irrelevant tweets)}
    \label{tab:states}
\end{table*}

\subsection{Learned States}

This section examines the states learned by the model and whether they align with course units or suggest different behavior units. Table~\ref{tab:states} summarizes the learned states with their topics and document type distributions (interpreted from $\phi$, $\theta_c$, and $\psi_c$). Most states are aligned with course units covering important course topics, such as learning analytics, data visualization, social networks, and prediction models. However, State 0 is where students do not participate in course discussion but post course-irrelevant tweets. State 3 is about hands-on practice of software tools across the course, and State 9 covers many side topics. 

Document types and their correlations with topics in each state also reveal interesting student behaviors (Table~\ref{tab:states}). Tweets tend to take a large proportion and goal notes a small proportion in every state due to their relative volumes. According to our examination of the data, blog posts are actively used for summarizing readings and tutorials, and tweets are used as a means of communicating with lecturers (e.g., State 5). ProSolo posts are most accessible to ProSolo users, so students use them to reveal their opinions and questions.
%Goal notes and blog posts are actively used in State 6 to discuss visualization tasks. Forum posts are popular in State 4 to discuss prediction models. Tweets are used much more than other types in State 5, discussing the tutorial lecture, and in State 9, discussing side topics.

\subsection{Students Following Goal Setters}

\begin{table}[t]
    \centering
    \small
    % \begin{tabularx}{\columnwidth}{lp{12mm}XXp{8mm}}
    \begin{tabularx}{\columnwidth}{lXXXX}
    \toprule
     & \multicolumn{4}{ c }{Social Connection} \\ \cmidrule{2-5}
     & GS \tiny{S1+S2} & GP \tiny{S3+S4} & GB \tiny{S5+S6} & NO \tiny{S7}   \\ \midrule
%    \# Students & 41 & 64 & 125 & 224  \\ 
%    \# Documents & 2487 & 8126 & 10096 & 7405 \\ 
    \# Time Points & 139 & 315 & 265 & 821 \\ \midrule
    \multicolumn{5}{ l }{\% Time Points}  \\ \midrule
    State 0 & \textbf{0.59$^{\star\star}$} & \textbf{0.75} & \textbf{0.75} & \textbf{0.71} \\
    State 1 & \textbf{0.17$^{\star}$} & \textbf{0.10} & \textbf{0.03} & \textbf{0.04} \\
    State 2 & 0.05 & 0.02 & 0.02 & 0.04 \\
    State 3 & \textbf{0.04$^{\star}$} & \textbf{0.00} & 0.01 & \textbf{0.01} \\
    State 4 & 0.01 & 0.02 & 0.03 & 0.06 \\
    State 5 & 0.05 & 0.03 & 0.06 & 0.05 \\
    State 6 & 0.05 & 0.02 & 0.02 & 0.02 \\
    State 7 & 0.03 & 0.01 & 0.03 & 0.02 \\
    State 8 & 0.00 & 0.03 & 0.02 & 0.02 \\
    State 9 & 0.01 & 0.04 & 0.03 & 0.04 \\ \bottomrule
    \end{tabularx}
    \caption{Proportion of time points students stay in each state depending on the social connection (each column sums to 1). ``$\star\star$'' and ``$\star$'' indicate that \textsf{GS} is significantly different from other categories in bold with $p < 0.01$ and $p < 0.05$, respectively, by Pearson's chi-square test. \textsf{GS}, \textsf{GP}, and \textsf{GB} each represent either ``has been following'' or ``started to follow'' a goal setter, a goal participant, and a goal bystander, respectively. \textsf{NO} means to follow no one.}
    \label{tab:states_by_status}
\end{table}

\begin{figure}[t]
    \centering
    \begin{subfigure}[t]{0.46\linewidth}
        \includegraphics[width=\linewidth]{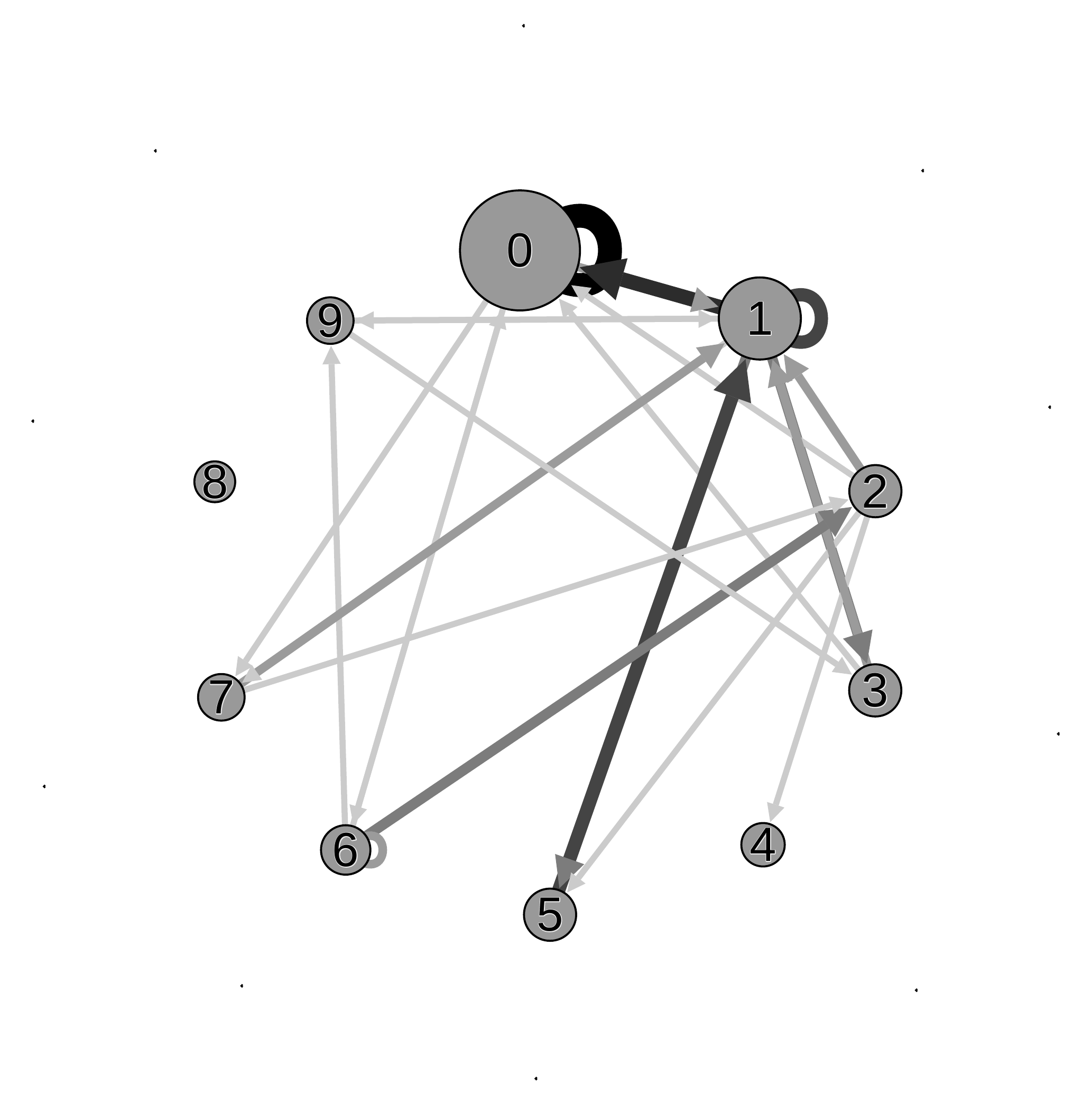}
        \caption{S1. Has been following a goal setter}
        \label{fig:transition_goalsetter}
    \end{subfigure}
    \begin{subfigure}[t]{0.46\linewidth}
        \includegraphics[width=\linewidth]{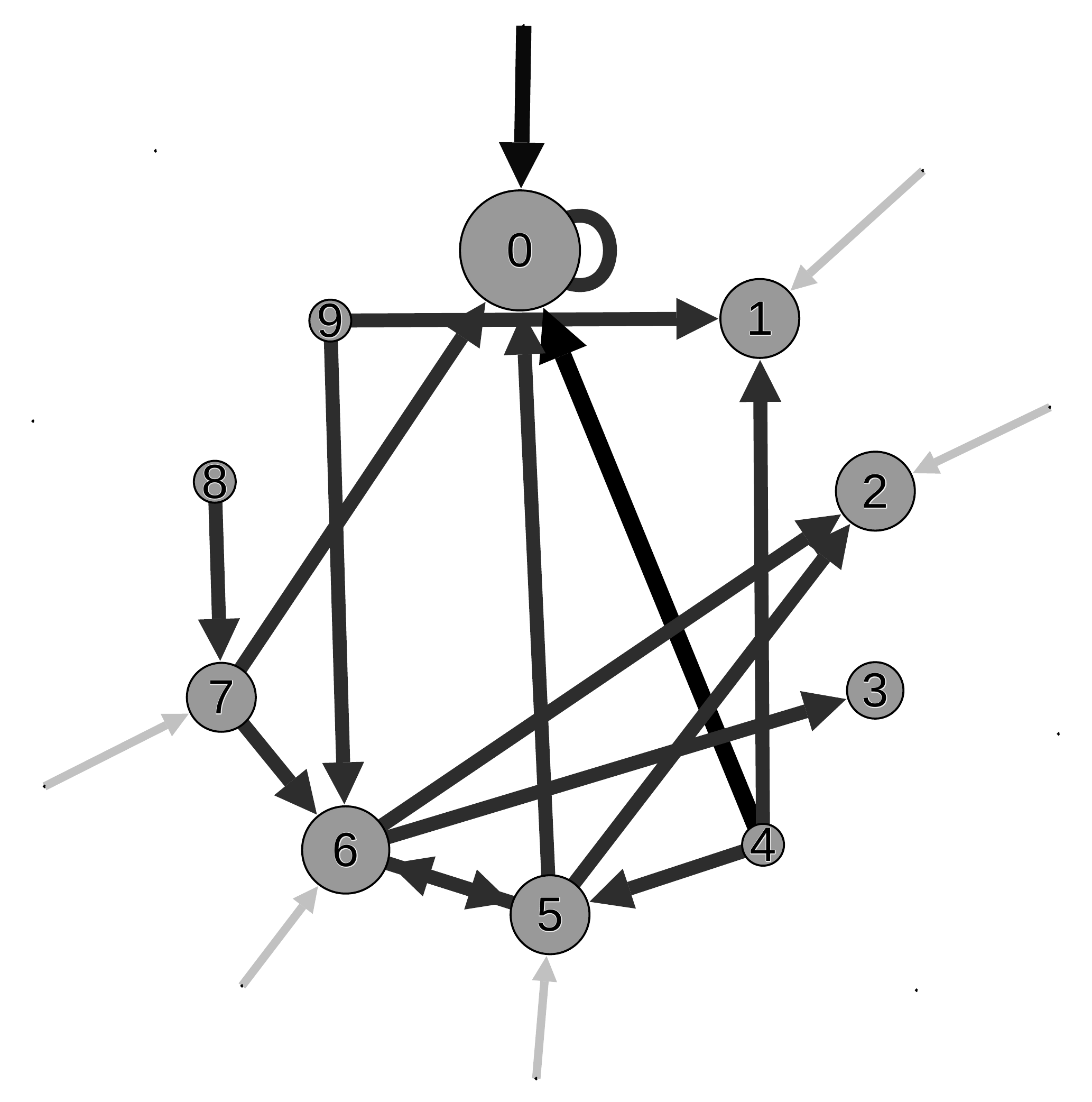}
        \caption{S2. Started to follow a goal setter}
        \label{fig:transition_startgoalsetter}
    \end{subfigure}\\
    \begin{subfigure}[t]{0.46\linewidth}
        \centering
        \includegraphics[width=\linewidth]{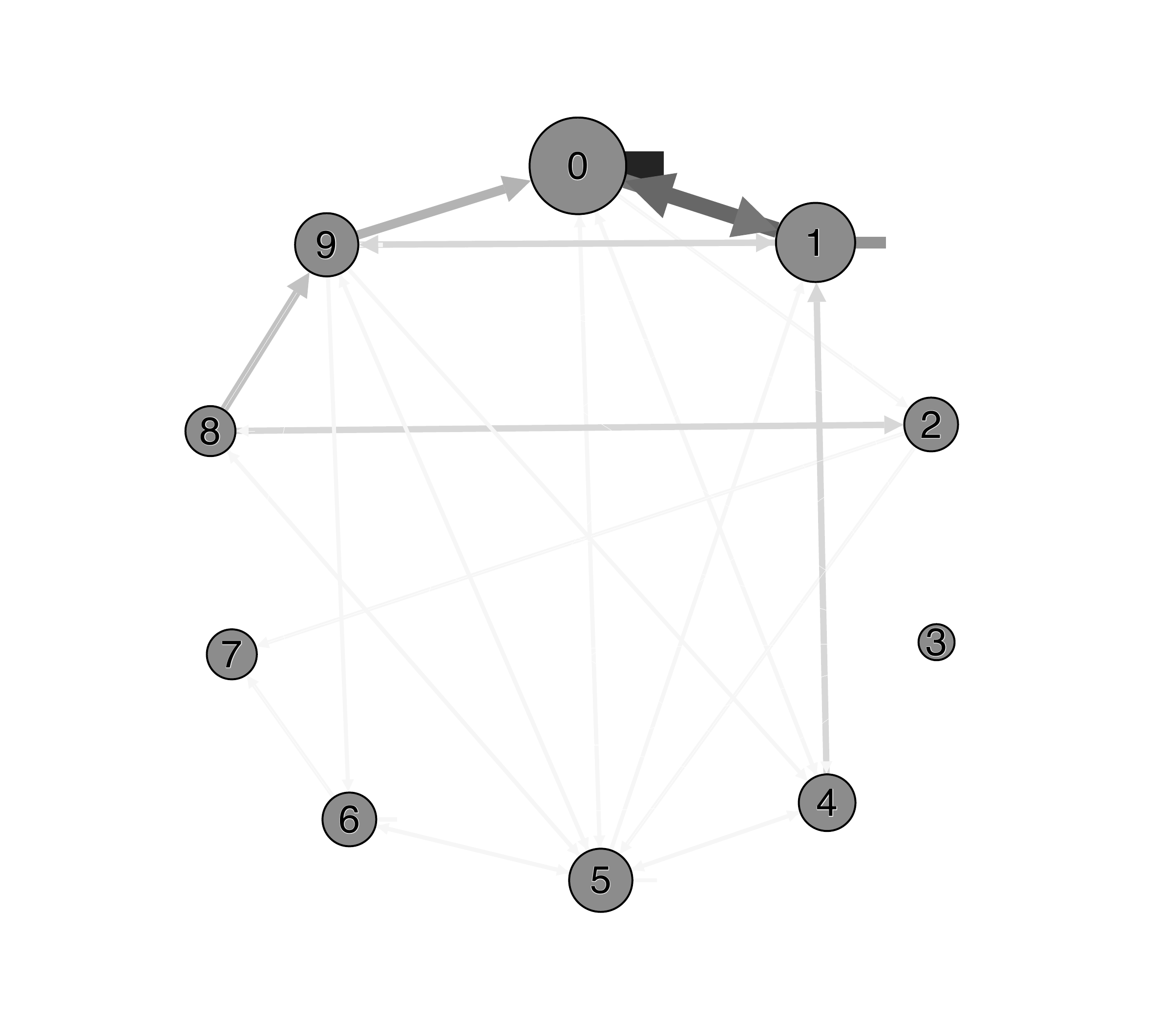}
        \caption{S3. Has been following a goal participant}
        \label{fig:transition_goalparticipant}
    \end{subfigure}
    \begin{subfigure}[t]{0.46\linewidth}
        \includegraphics[width=\linewidth]{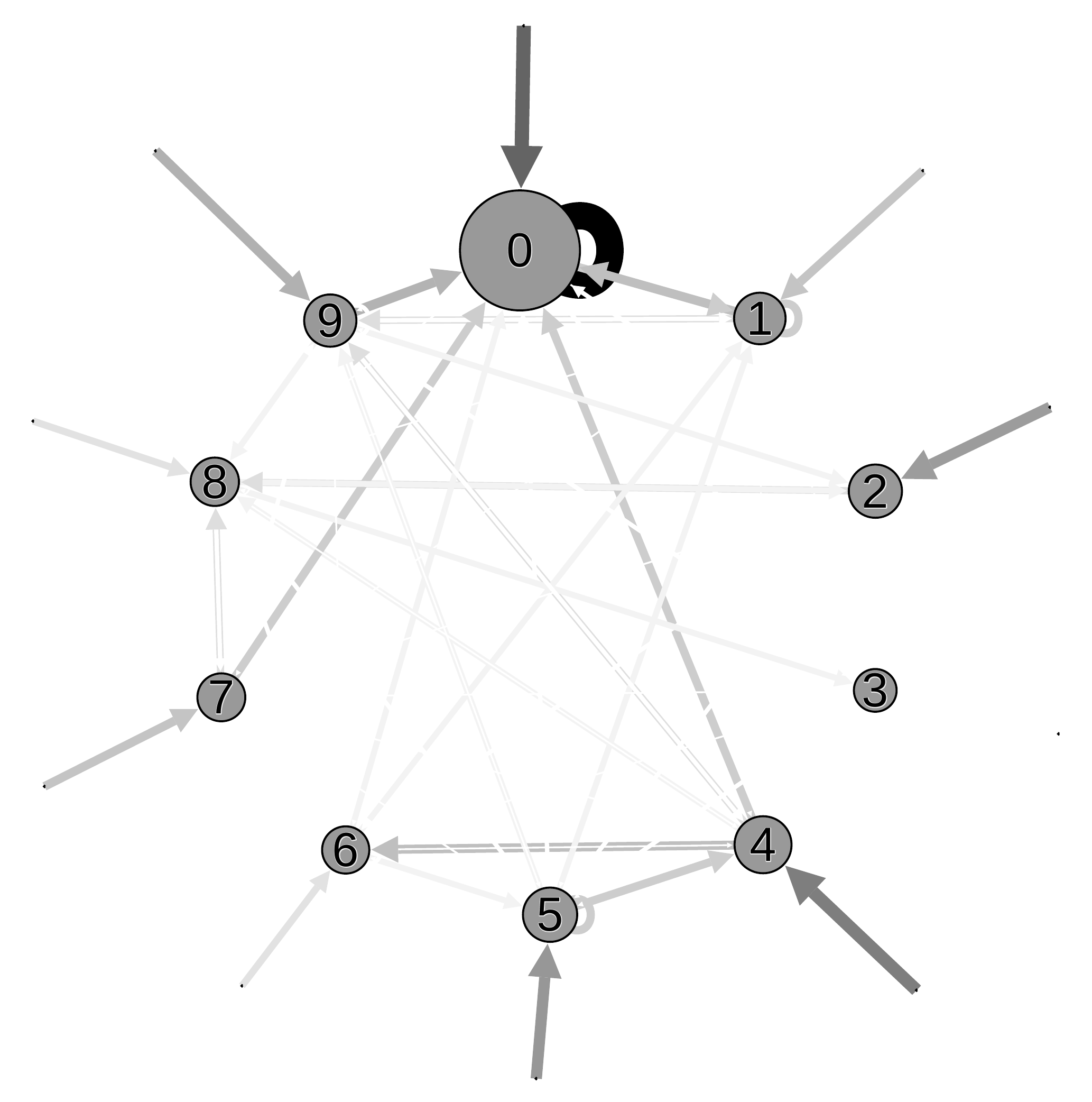}
        \caption{S7. Follows no one}
        \label{fig:transition_nofollowee}
    \end{subfigure}
    \caption{State transition patterns. Nodes are states whose size reflects the number of weeks students visit the states. Edges are transitions whose thickness and darkness reflect transition frequency. Edges without a source node represent the probability of being the first state in a learning path.}
    \label{fig:transition}
\end{figure}

We investigated the learning processes of the students who follow goal setters and their positive learning behaviors, based on the number of weeks students spent in each state (Table~\ref{tab:states_by_status}) and state transition patterns (Figure~\ref{fig:transition}). Note that in Table~\ref{tab:states_by_status}, for simplicity, ``has been following'' and ``started to follow'' categories are combined in each column.

\textbf{Twitter usage:} Students generally wrote irrelevant tweets (State 0) significantly more than course-related documents regardless of the social category. However, the students following goal setters spent noticeably fewer weeks in this state.

%\vspace{-3mm}
\textbf{Participation duration:} The topics of the states in which students stay reveal how long they persist in the course. The students following goal setters were more likely to discuss the material taught in the last week (State 1) than other students, that is, they were active in the last phase of the course. This phenomenon is consistent with the previous finding that social connection can lower attrition rates in MOOCs~\cite{Rose2014}.

% \vspace{-3mm}
\textbf{Activities of interest:} The number of weeks students spend in each state reflects the activities students are interested in. The students following goal setters were more active in hands-on practice (State 3) than other students. This state captures active learning, i.e., hands-on practice of software tools such as Tableau, Gephi, and Lightside across the course. Hands-on practice requires higher motivation than merely watching lectures, so these students might have been helped by observation of role models as discussed in the literature \cite{Schunk1985}. This trend would have not been as clear using predefined states based on course units~\cite{Coffrin2014}, which could not learn to distinguish lecture and practice activities.

% \vspace{-3mm}
\textbf{Study habits or challenges:} Transition patterns may reveal students' study habits or challenges. Figure~\ref{fig:transition_goalsetter} shows frequent transitions between three states (States 1, 3, and 5) that are associated with materials taught in different weeks. Such transitions may reflect the SRL strategy of activating and applying prior knowledge to the current situation~\cite{Pintrich2004}.

%\textbf{Short-term influence:} The transition patterns conditioned on ``start following'' give ideas about the short-term influence of making a social connection. Figure~\ref{fig:transition_startgoalsetter} shows that students who start following a goal setter have many transitions into visualization tasks (State 6) in the next time point. After than, these students remain there or move to the content of social capital as shown in Figure~\ref{fig:transition_goalsetter}.

These positive effects associated with following goal setters are not apparent with other social connection types, such as following goal participants or goal bystanders. The students who start to follow a goal setter (Figure~\ref{fig:transition_startgoalsetter}) begin to show the behaviors of those following a goal setter \mbox{(Figure~\ref{fig:transition_goalsetter})} changing from the behaviors of those following a goal participant (Figure~\ref{fig:transition_goalparticipant}). This indicates that ``who to follow'' is more important than simply following someone.

\subsection{Students With No Social Connection}

Students with no social connection were not passive users of ProSolo. Compared to students following goal setters, however, they were far more passive in writing documents and engaging with discussions. They also showed the following learning behavior (Table~\ref{tab:states_by_status}).

\textbf{Twitter usage:} Interestingly, the relative amount of time they spent in State 0 is similar to students who had social connections. They may be passive in using social media, or they may have not bothered to reveal their Twitter information. 

% \vspace{-3mm}
\textbf{Activities of interest:} While students with social connections had preferred states, these students were spread across all states quite evenly. They may have diverse interests but be indifferent about making connections with other people.

\chapter{Intervention for Support}\label{sec:support}

On the basis of the insights obtained from the previous component, the third component of our pipeline is to offer appropriate support, especially towards fostering beneficial social connections between students. We argue that a recommender system can serve this purpose, by presenting its potential positive impact as assessed on the corpus.

\section{Model}

Our recommender system aims to match qualified students (e.g., goal setters) to discussions so that they can interact with and benefit the discussants through discussions. Our model has two steps: relevance prediction and constraint filtering. The relevance prediction step learns the relevance between students and discussions using student- and discussion-related features that are potentially valuable in making recommendations. The learned relevance reflects students' preferences and tendencies, but may not reflect the ideal matches for fostering learning. The constraint filtering step thus combines the relevance scores with some constraints that foster interaction between qualified students and other students, and finalizes recommendations.

\subsection{Relevance Prediction}
Our algorithm extends the earlier model proposed by Yang et al.~\cite{yang2014question}, which is also designed to match students with discussions, by incorporating additional components related to students' qualifications. The relevance matrix between students and discussions is denoted as $R = \{r_{u,d}\}$ for every student $u$ and discussion $d$. $r_{u,d}$ is 1 if and only if student $u$ has participated in discussion $d$. As in the original model, our model exploits student features, discussion features, and implicit feedback. In addition, we add the following additional student features related to qualifications of interest.

% \vspace{-5mm}
\begin{itemize}
\item \textbf{Goal quality ($\lambda$):} A student's goal quality as defined in Section \ref{sec:research_context}. (2: goal setter, 1: goal participant, 0: goal bystander)
\item \textbf{Degree centrality ($\psi$):} The average of the authority and hub scores of a student's network. High centrality may serve as a hub for further social interaction.
%It has been calculated by taking average of authority and hub scores of a student in the social network in Prosolo. 
\end{itemize}

% \vspace{-3mm}
Taking these additional features into account, our new relevance prediction model can be formulated as:
\begin{equation}
 r_{u,d} = bias + \left( P_u + \phi_{u}\Phi + \theta_{u}\Theta+ \lambda_{u}\Lambda+ \psi_{u}\Psi+ \Gamma_{\gamma}\right)^{T} \times \nonumber \\
(Q_{d} + \delta_{d}\Delta + l_d L + \frac{1}{\sqrt{|U\left(d\right)|}}  \sum_{v \in U(d)} \varphi_{v}).
\end{equation}

% \vspace{-5mm}
$\phi_{u}$ and $\theta_{u}$ are the number of discussions $u$ has participated in and the number of discussions $u$ has initiated. $\Gamma_\gamma$ is a one-hot vector indicating the course week that $u$ registered for the course; this may be related to the student's motivation. $\delta_{d}$ and $l_d$ are the number of replies and the length of the content in $d$. $U(d)$ denotes the set of students participating in $d$, and $\varphi_{v}$ is the predicted preference $v$ of $u$. $P_u$ and $Q_d$ are the biases of $u$ and $d$. $\Phi$, $\Theta$, $\Lambda$, $\Psi$, $\Delta$, and $L$ are one-dimensional feature weights. Additional details regarding parameter estimation are available in Yang et al.'s paper~\cite{yang2014question}.

\subsection{Constraint Filtering}
It is important for students to receive relevant discussion recommendations, but it is also desirable to recommend students to discussions where the other discussants can benefit from the student's participation and thereby to optimize the overall community welfare. For this purpose, we use a max cost flow model to subject the relevance scores obtained in the previous step with the following constraints.

% \vspace{-5mm}
\begin{enumerate}
\item \textbf{Goal quality:} For every discussion $d$, at least one student $u$ to which we recommend $d$ should have a goal quality $G_{u}$ greater than some threshold $G$.

\item \textbf{Degree centrality:} For every discussion $d$, at least one student $u$ to which we recommend $d$ should have a centrality score $C_{u}$ greater than some threshold $C$. The rationale behind this constraint is that students with high centrality may serve as a hub through which more students are connected.

\item \textbf{Workload:} Qualified students should not be matched to too many discussions so that their workload is minimized.
\end{enumerate}

% \vspace{-5mm}
These constraints are formulated into the following optimization problem, where our goal is to compute $f_{u,d} \in \{0, 1\}$, an indicator of whether discussion $d$ is recommended to student $u$, that maximizes the objective function:
\begin{equation}
 \max \sum_{u,d} f_{u,d} \cdot r_{u,d} - \alpha \cdot \sum_{d}\sum_{u} \mathbbm{1}(G_{u}  \cdot f_{u,d} \ge G) (G_{u} -  G) \nonumber \\
 - \alpha \cdot \sum_{d}\sum_{u} \mathbbm{1}(C_{u}  \cdot f_{u,d} \ge C) (C_{u} -  C) ~~~~\text{s.t.} \nonumber \\
\forall d \in D, \exists u \in U, G_{u}  \cdot f_{u,d} \ge G \nonumber, \\
\forall d \in D, \exists u \in U, C_{u}  \cdot f_{u,d} \ge C, \label{eqn:obj}
\end{equation}

where $D$ and $U$ are the sets of discussions and students, respectively. $\mathbbm{1}$ is an indicator function. The second and third terms prevent qualified students from being assigned to too many discussions. The concept of a concave cost network for solving this optimization problem is detailed in Yang et al.'s paper~\cite{yang2014question}.

\section{Findings}

Since we have identified positive learning behaviors of students who follow goal setters, we may want to support students by fostering interaction with goal setters. Instead of recommending direct following relations, which are not supported by many learning platforms, we recommend discussions to qualified students so that they can interact with the discussants. We first assess the extent to which students are sensitive to qualified students prior to explicit intervention, and then present the potential added value of our recommendation model.

\subsection{Students' Awareness of Role Models}

Our first step is to assess whether students can identify effective role models in discussion activities (ProSolo posts), by measuring the impact of the information about students' qualifications on the prediction of discussion participation. This task is to infer links between students and discussions that we hid from an observed static snapshot of a network of discussion participation based on observable data. A measured positive impact here would indicate some sensitivity on the part of students to interact with qualified students naturally. 
%To simulate a realistic scenario, for each $r_{u,d} = 1$, we sampled five different discussions $d'$ such that $r_{u,d'}$ is not observed, and set $r_{u,d'}  = 0$. 
We train a predictive model of students' participation in discussions on two thirds of student-discussion pairs. We then predict the discussion participation of the remaining pairs.
%Note that the ground truth used to evaluate prediction results only contained the student participation history (whether the students participated in a discussion). 
Our evaluation metric is mean average precision (MAP).

We compared four configurations by varying the information about students' qualifications that is used as feature for relevance prediction. In particular, \textsf{CAMF} uses only basic features, such as the numbers of discussions each student initiated and participated in and each discussion's length, number of replies, and participants. \textsf{CAMF\_G} and \textsf{CAMF\_C} add information about goal quality and degree centrality, respectively, and \textsf{CAMF\_GC} adds both. The evaluation was conducted as a link prediction task, based on the relevance scores predicted in the relevance prediction step. Students' qualification information did not improve link prediction accuracy (Table~\ref{tbl:diagnosis}). This means that students are not proactively sensitive to peers' qualifications while participating in discussions, which supports our view that explicit recommendation could be valuable for encouraging students to interact with qualified peers through discussions.

\subsection{Recommendation Quality}
The recommendation of discussions should be consistent with both the relevance between students and discussions (the relevance prediction step) and constraints for beneficial social connection (the constraint filtering step). To this end, we evaluated recommendation quality on Overall Community Benefit (OB) (Equation \ref{eqn:obj} without the constraints): the relevance of our recommendations penalized by the burden on the students induced by the recommendations. The higher OB the better. 

We tested three configurations by varying the constraints incorporated into the constraint filtering step. \textsf{MCCF\_G} requires that every discussion have at least one goal participant or goal setter. \textsf{MCCF\_C} requires that every discussion have at least one student whose degree centrality is higher than 0.1. \textsf{MCCF\_GC} requires both. In addition, the following configurations were tested as baseline without incorporation into the model. \textsf{GoalPart} filters goal participants or goal setters after making recommendations based on predicted relevance. Similarly, \textsf{HighCent} filters students with degree centrality higher than 0.1. \textsf{GoalPart\_HighCent} filters goal participants or goal setters with degree centrality higher than 0.1. Incorporating the constraints about students' goal quality and degree centrality into the model (\textsf{MCCF\_G}, \textsf{MCCF\_C}, and \textsf{MCCF\_GC}) achieved higher OB than the simple filtering approaches (Table~\ref{tbl:community_benefit}). That is, our algorithm effectively matches qualified models to relevant discussions in such a way that students in every discussion can interact with qualified models while balancing the load of the models.

\begin{table}[tbp]
\small
\centering
\begin{tabularx}{\columnwidth}{ p{20mm}X p{20mm}X }
\toprule
Configuration & MAP & Configuration & MAP\\\midrule
CAMF & 0.465 & CAMF\_C & 0.455\\
CAMF\_G & 0.438 & CAMF\_GC & 0.439\\
\bottomrule
\end{tabularx}
\caption{MAP for link prediction.}
\label{tbl:diagnosis}
\end{table}

\begin{table}[tbp]
\small
\centering
\begin{tabularx}{\columnwidth}{ p{26mm}X p{20mm}X }
\toprule
Configuration & OB & Configuration & OB\\\midrule
GoalPart & 1.888 & MCCF\_G & 3.683\\
HighCent & 1.943 & MCCF\_C & 3.770\\
GoalPart\_HighCent & 1.873 & MCCF\_GC & 3.656\\
\bottomrule
\end{tabularx}
\caption{Overall Community Benefit for recommendation.}
\label{tbl:community_benefit}
\end{table}

\chapter{Discussion}

According to our learning process analysis, students benefit from social connections with effective goal setters through ProSolo's follower-followee functionality. They stay longer in the course, engage in hands-on practices, and link materials across the course. This supports the view that goal-setting behavior is a useful qualification for potential role models. According to the discussion participation prediction task, explicit intervention is important for helping students be aware of qualified students and interact with them via discussions. Therefore, we incorporated the information about students' qualifications into our recommendation model as constraints, successfully matching qualified learning partners to relevant discussions.

This work started from the need for expediting data analysis and analysis-informed support in social learning where students interact with one another via various social media in order to pursue their own learning goals. This expedition builds on DiscourseDB, data infrastructure for complex interaction data from heterogeneous platforms. We proposed a probabilistic graphical model to analyze students' learning processes depending on the state of their social connections, and proposed a recommender system that can improve student support on the basis of the insights obtained from the analysis. This pipeline arguably should allow us to apply the techniques to different learning communities with little effort.

Goal-setting behavior is an important practice in SRL and is known to be difficult for students, so an analysis towards improvement of this skill is arguably valuable. Nevertheless, in this study we have not examined how this behavior influences the domain learning of students. This is due both to the limited data size for our first trial to use ProSolo in MOOCs as well as a lack of learning gain measures. However, the modeling techniques proposed in this paper can readily be applied to other data sets if the requisite data become available. We are also interested in investigating different SRL strategies besides goal-setting in social learning, and how social interaction influences the SRL behaviors of the students. Ultimately, the real value of the work will be demonstrated not with a corpus analysis, as for our proposed recommendation approach, but with an intervention study in a real MOOC. We are working towards incorporating this approach in a planned rerun of DALMOOC.

\begin{acknowledgments}
This research was supported by the National Science Foundation under grants ACI-1443068 and IIS-1320064, and by the Naval Research Laboratory and Google.
\end{acknowledgments}

%\appendix
%\include{appendix}

\backmatter

%\renewcommand{\baselinestretch}{1.0}\normalsize

% By default \bibsection is \chapter*, but we really want this to show
% up in the table of contents and pdf bookmarks.

%\newcommand{\bibpreamble}{This text goes between the ``Bibliography''
%  header and the actual list of references}
\bibliographystyle{plainnat}
\bibliography{sigproc.bib} %your bib file

\end{document}